\documentclass{article}
\usepackage{amssymb}
\usepackage{amsmath}
\usepackage{graphicx}
\usepackage{cite}

\begin{document}

\title{Birth and metamorphoses of resonances in the driven van der Pol oscillator}
\author{Jan Kyzio\l , Andrzej Okni\'nski \\
Politechnika \'Swi\c{e}tokrzyska, Al. 1000-lecia PP 7, \\
25-314 Kielce, Poland}
\maketitle

\begin{abstract}
The dynamics of the driven van der Pol oscillator are investigated. We study birth and metamorphoses 
of $1:1$ and $1:3$ resonances  within the formalism of differential properties of amplitude-frequency 
response implicit functions. 
\end{abstract}

\section{Introduction}
\label{intro}

We study the forced van der Pol equation in nondimensional form
\begin{equation}
\dfrac{d^{2}z}{d\tau ^{2}}-\mu \left( 1-z^{2}\right) \dfrac{dz}{d\tau }%
+z=F\cos \left( \Omega \tau \right) ,  \label{vdP}
\end{equation}%
where $\tau $ is a nondimensional time and $\mu $ is a positive damping 
parameter.

Equation (\ref{vdP}) was introduced by van der Pol to describe vacuum-tube
oscillations \cite{vdPol1927} (see also \cite{vdPol1926} for the non-forced
equation). Investigations of vacuum-tube oscillations led to the early discovery
of deterministic chaos, noticed by van der Pol and van der Mark in an 
electrical circuit described by Eq. (\ref{vdP}) \cite{vdPolvdMark1927}. Equation (\ref{vdP}%
)\ was investigated further by Cartwright and Littlewood \cite%
{Cartwright1945} to confirm the presence of chaotic dynamics. 

The van der Pol oscillator displays several generic nonlinear phenomena; see
for global bifurcation phenomena Refs. \cite%
{Holmes1978,Parlitz1987,Mettin1993,Guckenheimer2003a,Guckenheimer2003b,Jordan2007}.

The first application of the van der Pol oscillators outside electrical
engineering, carried out by van der Pol and van der Mark, was to describe
the heartbeat \cite{vdPolvdMark1928,Pikovsky2001}. Then, the van der Pol equation has
been used to model self-sustained oscillations in physics, mechanics,
engineering, and biology.

In this paper, we study birth and metamorphoses of $1:1$ and $1:3$ resonances 
in the framework of the differential properties of amplitude-frequency 
response implicit functions \cite{Gibson1998,Kyziol2022},  continuing our previous work on van der Pol-Duffing 
equation \cite{Kyziol2015,Kyziol2019}.

\section{Nonlinear resonances via KBM method}
\label{KBM}

We apply the Krylov-Bogoliubov-Mitropolsky (KBM) perturbation approach \cite%
{Nayfeh2011} to Eq. (\ref{vdP}). We assume a resonance in form 
\begin{equation}
z_{\alpha }\left( \tau \right) =A\cos \left( \alpha \Omega \tau +\varphi
\right) +\varepsilon z_{1}\left( A,\varphi ,\tau \right) +\ldots ,  \label{z}
\end{equation}
where the amplitude $A$ and frequency $\Omega $ fulfil the
amplitude-frequency response equation%
\begin{equation}
\mathcal{F}_{\alpha }\left( A,\Omega ,\mu ,F\right) =0.  \label{a-f_1}
\end{equation}%
In what follows we consider $\alpha =1$ and $\alpha =\frac{1}{3}$ for $1:1$
and $1:3$ resonances, respectively. In the case of primary resonance, we
have 
\begin{equation}
\begin{array}{lll}
\mathcal{F}_{1:1}\left( A,\Omega ,\mu ,F\right)  & = & A^{2}\left( \mu
^{2}\Omega ^{2}\left( 1-\tfrac{1}{4}A^{2}\right) ^{2}+\left( 1+\tfrac{3}{4}%
\lambda A^{2}-\Omega ^{2}\right) ^{2}\right) -F^{2},%
\end{array}
\label{1:1}
\end{equation}
while for the $1:3$ resonance we get
\begin{equation}
\begin{array}{lll}
\mathcal{F}_{1:3}\left( A,\Omega ,\mu ,F\right)  & =\medskip  & \Omega
^{4}\mu ^{4}\left( A^{2}-4+\tfrac{2F^{2}}{\left( \Omega ^{2}-1\right) ^{2}}%
\right) ^{2}+\Omega ^{2}\mu ^{2}\left( \tfrac{4}{3}\Omega ^{2}-12\right) ^{2}
\\ 
&  & -\tfrac{\Omega ^{4}A^{2}F^{2}\mu ^{4}}{\left( \Omega ^{2}-1\right) ^{2}}%
\quad \left( \Omega \neq 1\right) .%
\end{array}
\label{1:3}
\end{equation}

\section{Singular and critical points of the amplitude profiles}
\label{general}

Singular points of an implicit function $L\left( A,\Omega ,\underline{c}%
\right) =0$, where $\underline{c}=c_{1},c_{2},\ldots $ are parameters, are
solutions of equations 
\begin{subequations}
\label{SINGULAR}
\begin{eqnarray}
L\left( A,\Omega ,\underline{c}\right)  &=&0,  \label{S1} \\
\frac{\partial L\left( A,\Omega ,\underline{c}\right) }{\partial \Omega }
&=&0,  \label{S2} \\
\frac{\partial L\left( A,\Omega ,\underline{c}\right) }{\partial A} &=&0.
\label{S3}
\end{eqnarray}
\end{subequations}
Solutions of Eqs. (\ref{SINGULAR}), if exist, are of form $A=A_{\ast }$, $\Omega
=\Omega _{\ast }$, $\underline{c}=\underline{c}_{\ast }$. 

It follows from Eqs. (\ref{S2}) and (\ref{S3}), that in the neighborhood of a singular point $ 
\left( A_{\ast },\Omega _{\ast }\right) $ of the function $L\left( A,\Omega ,%
\underline{c}_{\ast }\right) =0$, neither of the functions $\Omega =f\left(
A\right) $ nor $A=g\left( \Omega \right) $ is single-valued. 

Moreover, equations (\ref{S1}) and (\ref{S2}) are conditions for extremas of
the function $A=g\left( \Omega \right) $, while equations (\ref{S1}), (\ref%
{S3}) are conditions for vertical tangencies of the function $A=g\left(
\Omega \right) $ or extremas of the function $\Omega =f\left( A\right) $.

\subsection{The case of resonance $1:1$}
To compute singular points of the amplitude-frequency implicit function for the $
1:1$ resonance, we substitute in (\ref{SINGULAR}) $L\left( A,\Omega ,%
\underline{c}\right) =\mathcal{F}_{1:1}\left( A,\Omega ,\mu ,F\right) $.

Solutions of Eqs. (\ref{SING1:1})

\begin{subequations}
\label{SING1:1}
\begin{eqnarray}
\mathcal{F}_{1:1}\left( A,\Omega ,\mu ,F\right)  &=&0  \label{a} \\
\frac{\partial \mathcal{F}_{1:1}\left( A,\Omega ,\mu ,F\right) }{\partial A}
&=&0  \label{b} \\
\frac{\partial \mathcal{F}_{1:1}\left( A,\Omega ,\mu ,F\right) }{\partial A}
&=&0  \label{c}
\end{eqnarray}%
\end{subequations}
read 

\begin{equation}
A=2,\ \Omega =1,\ F=0,\quad \left( \mu \neq 0\right)   \label{1:1_1}
\end{equation}
and%
\begin{equation}
\hspace{-6pt}\left. 
\begin{array}{l}
5\mu ^{2}A^{6}-44\mu ^{2}A^{4}+\left( -192+112\mu ^{2}\right)
A^{2}+256-64\mu ^{2}=0 \\ 
8\Omega ^{2}+16\mu ^{2}-8\mu ^{2}A^{2}+\mu ^{2}A^{4}-32=0 \\ 
\left( \mu ^{4}-5\mu ^{2}\right) A^{4}+\left( -120+38\mu ^{2}-8\mu
^{4}\right) A^{2}-104\mu ^{2}+160+16\mu =5^{3}F^{2}%
\end{array}%
\right\}   \label{1:1_2}
\end{equation}
Solution (\ref{1:1_1}) corresponds to an isolated point, while solution (\ref{1:1_2}) represents 
self-intersections. 
\begin{figure}[h!]
\center
\includegraphics[width=10.5cm, height=7cm]{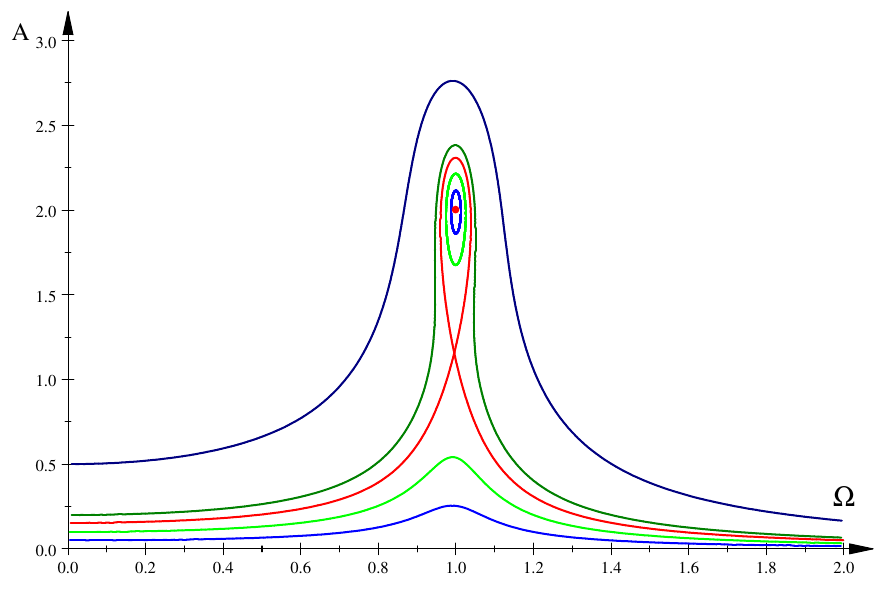}
\caption{Amplitude-frequency response functions $\mathcal{F}_{1:1}\left( A,\Omega
,\mu ,F\right) =0$. $\mu =0.2$; $F=0.05, 0.10, 0.153\,618 , 0.20, 0.50$
(LightBlue, LightGreen, LightRed, Green, Blue). LightRed dot marks an
isolated point.}
\label{Fa}
\end{figure}

Moreover, solutions of Eqs. (\ref{VT1:1}) for vertical tangencies 
\begin{subequations}
\label{VT1:1}
\begin{eqnarray}
\mathcal{F}_{1:1}\left( A,\Omega ,\mu ,F\right)  &=&0  \label{d} \\
\frac{\partial \mathcal{F}_{1:1}\left( A,\Omega ,\mu ,F\right) }{\partial A}
&=&0  \label{e}
\end{eqnarray}
\end{subequations}
are
\begin{equation}
\hspace{-5pt}\left. 
\begin{array}{l}
2\mu ^{4}A^{12}+\left( -3F^{2}\mu ^{4}-16\mu ^{4}\right) A^{10}+\left(
28F^{2}\mu ^{4}+32\mu ^{4}\right) A^{8} \\ 
+\left( -80F^{2}\mu ^{4}+32F^{2}\mu ^{2}\right) A^{6}+\left( 64F^{2}\mu
^{4}-128F^{2}\mu ^{2}\right) A^{4}+128F^{4}=0 \\ 
2A^{6}\mu ^{2}-\left( 3F^{2}\mu ^{2}+8\mu ^{2}\right) A^{4}+16\mu
^{2}F^{2}A^{2}-16F^{2}\mu ^{2}+32F^{2}=16F^{2}\Omega ^{2}%
\end{array}%
\right\}   \label{1:1_3}
\end{equation}

\begin{figure}[h!]
\center
\includegraphics[width=10.5cm, height=7cm]{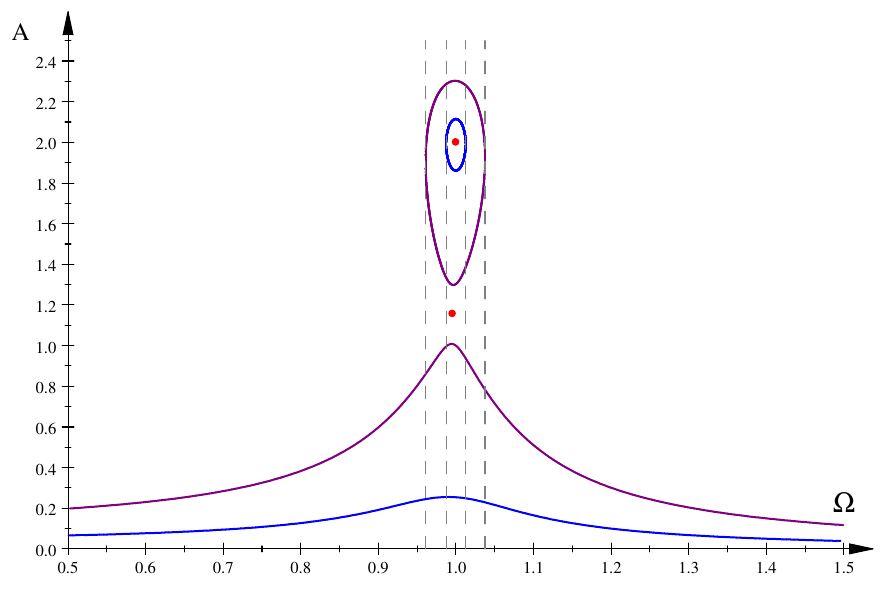}
\caption{Amplitude-frequency response functions $\mathcal{F}_{1:1}\left( A,\Omega
,\mu ,F\right) =0$. $\mu =0.2$; $F=0.05, 0.15$ (LightBlue, Purple). LightRed
dots mark singular points. Dashed lines show vertical tangencies. 
}
\label{Fb}
\end{figure}

\subsection{The case of resonance $1:3$}
Since the implicit function $\mathcal{F}_{1:3}\left( A,\Omega ,\mu ,F\right)
=0$ depends on $A^{2}$, $\Omega ^{2}$, $\mu ^{2}$, and $F^{2}$ only we write
Eq. (\ref{1:3}) in a simpler form
\begin{equation}
\hspace{-4pt}L_{1:3}\left( Y,X,m,f\right) =X^{2}m^{2}\left( Y-4+\tfrac{2f}{%
\left( X-1\right) ^{2}}\right) ^{2}+Xm\left( \tfrac{4}{3}X-12\right) ^{2}-%
\tfrac{X^{2}Yfm^{2}}{\left( X-1\right) ^{2}},  \label{L}
\end{equation}
where $Y=A^{2}$, $X=\Omega ^{2}$, $m=\mu ^{2}$, $f=F^{2}$, and $%
L_{1:3}\left( A^{2},\Omega ^{2},\mu ^{2},F^{2}\right) =\mathcal{F}%
_{1:3}\left( A,\Omega ,\mu ,F\right) $.

Singular points of the implicit function $L_{1:3}\left( Y,X,m,f\right) =0$
are solutions of the following equations
\begin{subequations}
\label{SING}
\begin{eqnarray}
L_{1:3}\left( Y,X,m,f\right)  &=&0,  \label{s1} \\
\frac{\partial L_{1:3}\left( Y,X,m,f\right) }{\partial X} &=&0,  \label{s2}
\\
\frac{\partial L_{1:3}\left( Y,X,m,f\right) }{\partial Y} &=&0.  \label{s3}
\end{eqnarray}
\end{subequations}
We note here that there are two kinds of singular points in these systems:
isolated points and self-intersections. The isolated points are signatures of 
the birth of a new branch of solution, while self-intersections correspond to
rupture of existing branches.

\begin{figure}[h!]
\center
\includegraphics[width=10.5cm, height=7cm]{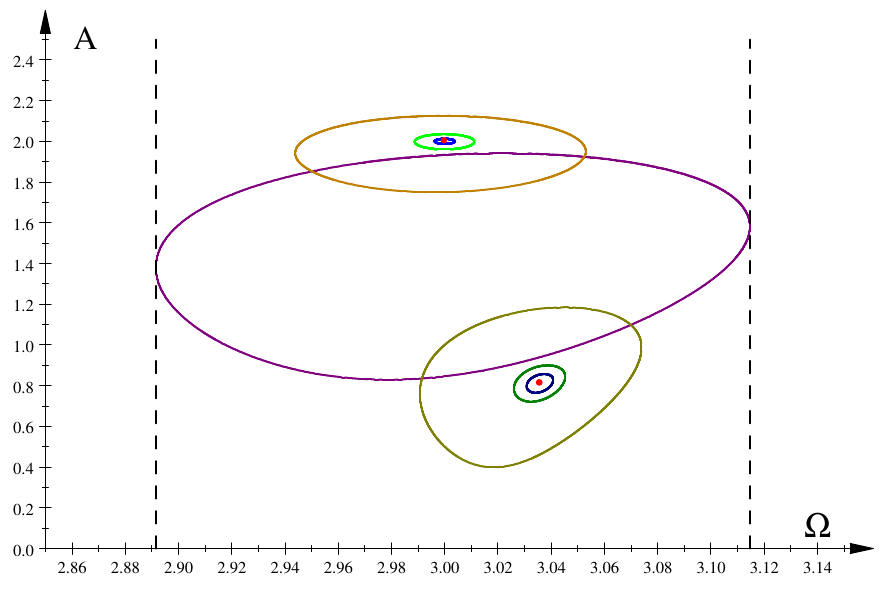}
\caption{Amplitude-frequency response functions $L_{1:3}\left( Y,X,m,f\right) =0$. $%
\mu =0.2$; $F=0.2$, $0.6$, $3$, $8.7$, $12$, $12.25$, $12.26$ (LightBlue,
LightGreen, Sienna, Purple, Brown, Green, Blue). LightRed dots mark isolated
points. Dashed lines show vertical tangencies. }. 
\label{Fc}
\end{figure}

Moreover, vertical tangencies are computed from 
\begin{subequations}
\label{VT}
\begin{eqnarray}
L_{1:3}\left( Y,X,m,f\right)  &=&0,  \label{v1} \\
\frac{\partial L_{1:3}\left( Y,X,m,f\right) }{\partial Y} &=&0.  \label{v2}
\end{eqnarray}
\end{subequations}
Vertical tangencies show the range of a resonance, while self-intersections
correspond to rupture of existing branches.

We solve Eqs. (\ref{SING}) to get the following singular solutions 
\begin{subequations}
\begin{equation}
X=9,Y=4,f=0,\text{\quad }\left( m\text{ arbitrary}\right)  \label{vdP1}
\end{equation}
which is an isolated point, and another solution%
\begin{equation}
\left\{ 
\begin{array}{l}
Y=\tfrac{-5X^{3}+73X^{2}+12mX^{2}-243X-81}{3mX^{2}},\medskip \\ 
f=4\tfrac{-4354X^{3}+11\,732X^{2}+17%
\,010X+3969+315X^{4}+108X^{4}m-360X^{3}m-324mX^{2}}{315mX^{2}},\medskip \\ 
175X^{5}+\left( -216m-3535\right) X^{4}+\left( 720m+22\,498\right) X^{3} \\ 
\qquad \hspace{10pt}+\left( 648m-40\,194\right) X^{2}-31\,185X-5103=0,%
\end{array}%
\right.  \label{vdP2}
\end{equation}
\end{subequations}
with arbitrary $m$.

Condition that discriminant of equation for $X$ in (\ref{vdP2}) vanishes
yields $m=-28$, $m=0$, and 

\begin{gather}
24\,564\,384m^{4}+1317\,273\,111m^{3}+104\,720\,016\,996m^{2}  \label{dis} \\
-1475\,275\,114\,404m-178\,499\,175\,680=0.  \notag
\end{gather}
A positive solution of Eq. (\ref{dis}) is $m_{cr}=12.\,009\,279$.
Accordingly, equation for $X$ in Eq. (\ref{vdP2}) has one positive root for $%
m<m_{cr}$, corresponding to an isolated point, and three positive roots for $%
m>m_{cr}$, corresponding to two isolated points and one self-intersection.
\bigskip

To compute vertical tangencies, we solve  equations (\ref{VT}) obtaining 
\begin{subequations}
\label{vvdP}
\begin{gather}
Y=\tfrac{8X^{2}-16X-3f+8}{2\left( 1-X\right) ^{2}},  \label{vY} \\
64X^{6}-1408X^{5}+10\,176X^{4}+\left( -27\,904-144mf\right) X^{3}  \label{vX}
\\
+\left( 35\,776+288mf\right) X^{2}+\left( 63mf^{2}-21\,888-144mf\right)
X+5184=0.  \notag
\end{gather}
\end{subequations}
In the case when two vertical tangencies overlap, we have a singular point --
an isolated point. This happens when Eq. (\ref{vX}) has a double root, i.e.
when a discriminant, $D\left( m,f\right) $, of Eq. (\ref{vX}) vanishes.
Computing the discriminant, we get 

\begin{equation}
D=\left( 
\begin{array}{l}
987\,614\,208m^{4}f^{4}-4514\,807\,808f^{3}m^{4}+5159\,780\,352f^{2}m^{4} \\ 
+27\,\allowbreak 205\,113\,600m^{3}f^{4}-30\,\allowbreak
059\,237\,376m^{3}f^{3}-1056\,\allowbreak 502\,185\,984f^{2}m^{3} \\ 
+761\,\allowbreak
014\,517\,760fm^{3}-472\,696\,875f^{5}m^{2}+96\,\allowbreak
981\,192\,000m^{2}f^{4} \\ 
-3830\,\allowbreak 568\,505\,344m^{2}f^{3}-11\,576\,\allowbreak
272\,748\,544m^{2}f^{2}-60\,542\,\allowbreak 932\,746\,240fm^{2} \\ 
-22\,265\,\allowbreak 110\,462\,464m^{2}-55\,\allowbreak
535\,398\,912f^{3}m-11\,861\,\allowbreak 638\,512\,640mf^{2} \\ 
+2635\,718\,\allowbreak 350\,340\,096mf+2216\,615\,\allowbreak
441\,596\,416m+377\,132\,\allowbreak 488\,327\,168f \\ 
-55\,169\,095\,\allowbreak 435\,288\,576%
\end{array}%
\right)  \label{D}
\end{equation}

\begin{figure}[ht!]
\center
\includegraphics[width=10.5cm, height=7cm]{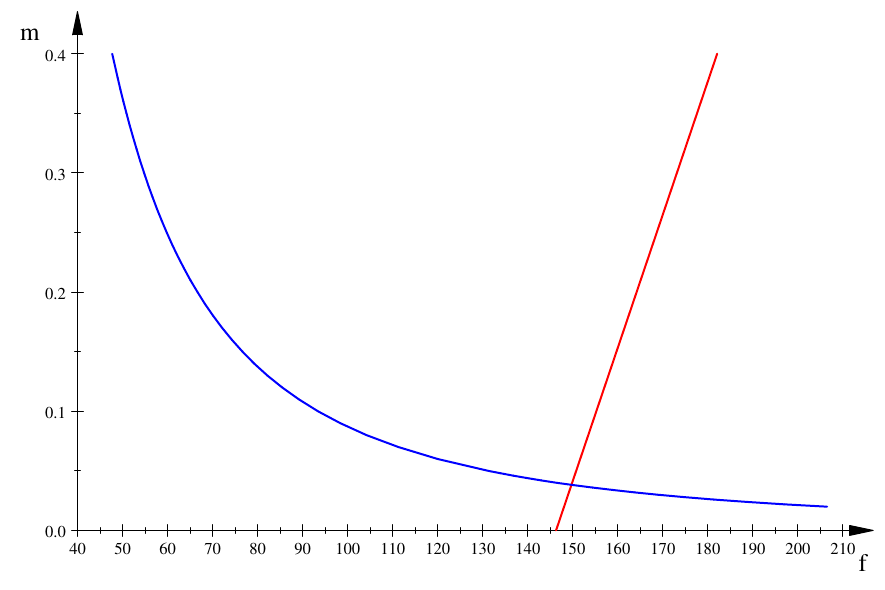}
\caption{Manifold of singular (isolated) points. Physical points lie on the LightRed
interval. .}
\label{Fd}
\end{figure}

The plot of $D\left( m,f\right) =0$ is shown in Fig. \ref{Fd}. Figure \ref
{Fd} shows values of parameters $f$, $m$ for which the function $L\left(
Y,X,m,f\right) =0$ has singular (isolated) points (red and blue curves). Values of  $f$ and $m$ that
lead to physical values for $X$ and $Y$ are on the red interval.
\newpage

\section{Computational results}
\label{Computations}

In this Section, we numerically solve Eq. (\ref{vdP}) and compare the results with analytical predictions. 
\subsection{The case of resonance $1:1$}

The solution (\ref{1:1_1}) of Eqs. (\ref{SING1:1}) describes a well-known
limit cycle of the unforced van der Pol equation \cite{Jordan2007}.
This solution is a good approximation to the limit cycle for small $\mu$, while for large values of $\mu$ 
a good approximation can be found in \cite{Dorodnicyn1947,Zonneveld1966}. 

It follows from Figs. \ref{Fa} and \ref{Fb} that for $F<0.153\,618$ the
amplitude-frequency response function for the $1:1$ resonance consists of
two disjoint pieces, where, as we were able to determine, only the upper
part is stable.

We have solved Eqs. (\ref{VT1:1}) for vertical tangencies setting $\mu =0.2$ and $%
F=0.05$, and $F=0.15$. We have obtained $\Omega _{1}=0.987\,395$, $%
A_{1}=1.\,991\,838$ and $\Omega _{2}=1.\,012\,447$, $A_{2}=1.\,992\,244$ for 
$F=0.05$ and $\Omega _{1}=0.960\,887$, $A_{1}=1.\,905\,297$ and $\Omega
_{2}=1.\,037\,503$, $A_{2}=1.\,\allowbreak 921\,864$ for $F=0.15$.

In Fig. \ref{Fb}, the computed vertical tangencies are shown. The
corresponding upper parts of the amplitude profiles are in the intervals $
\left( \Omega _{1},\Omega _{2}\right) $. 

\begin{figure}[h!]
\center
\includegraphics[width=10.5cm, height=7cm]{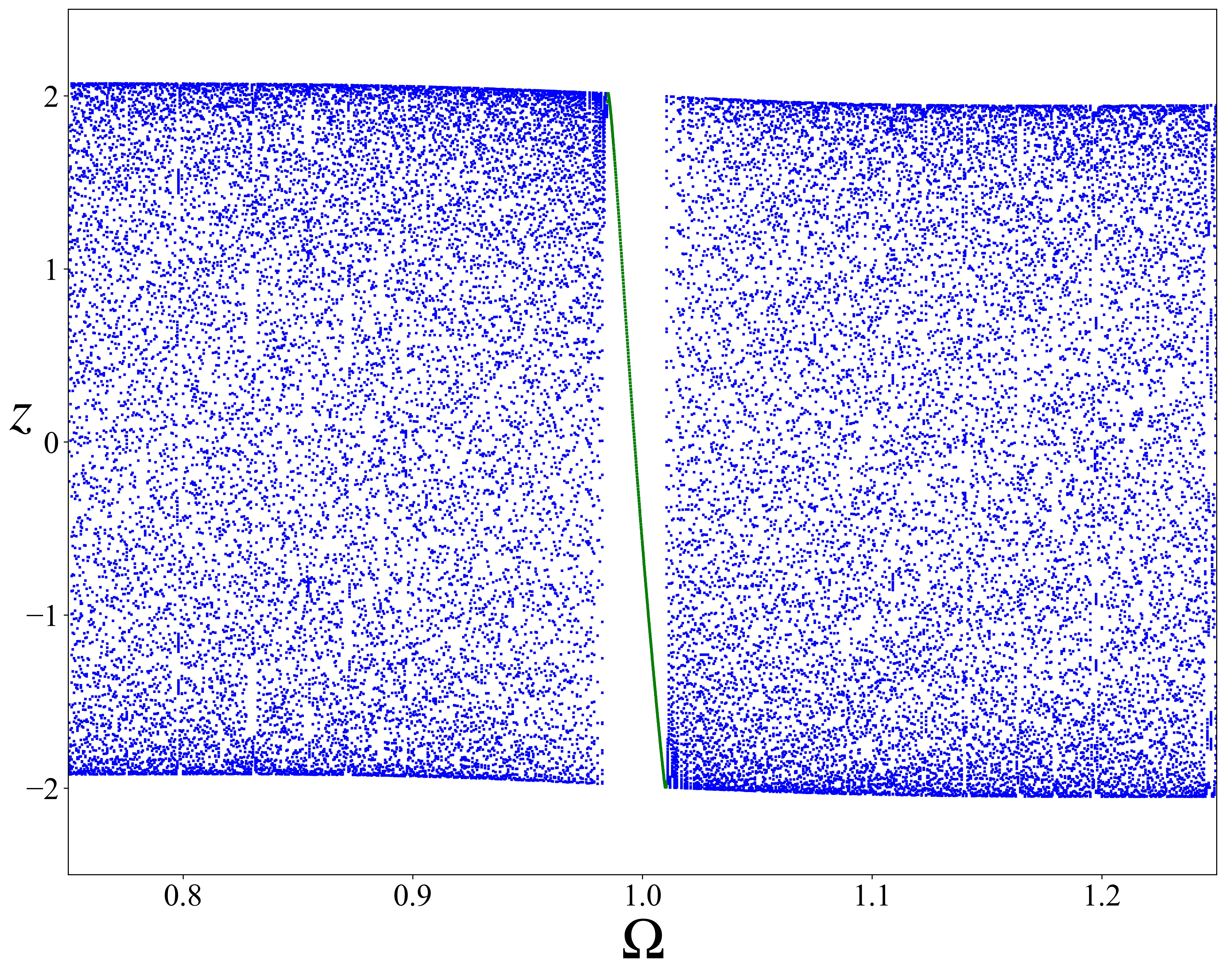}
\includegraphics[width=10.5cm, height=7cm]{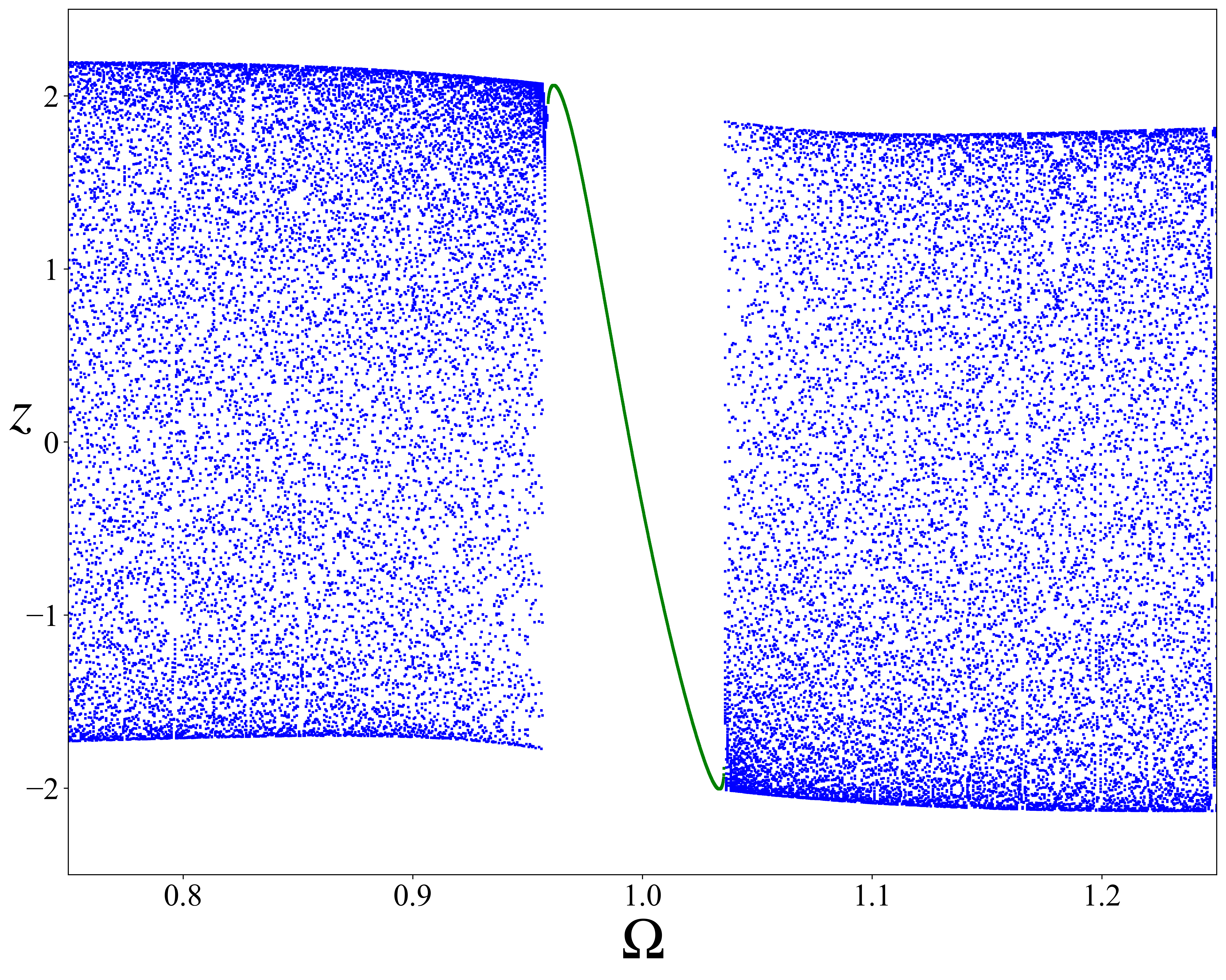}
\caption{The birth and growth of $1:1$ resonance (Green) -- numerical solutions of
Eq. (\ref{vdP}). $\mu =0.2; F=0.05$ (top), $0.15$ (bottom). Chaotic dynamics
is Blue.}
\label{F1a}
\end{figure}

We have solved Eq. (\ref{vdP}) numerically for $\mu =0.2$ and $F=0.05$, and $%
F=015$; see Fig. \ref{F1a} for the corresponding bifurcation diagrams. We
have determined that for $F=0.05$, the $1:1$ resonance exists in the interval 
$\Omega \in \left( 0.9848,1.0104\right) $, while for $F=0.15$ the primary
resonance is in the interval $\Omega \in \left( 0.9589,1.0355\right) $. We
notice a good agreement between numerical and analytical results.

\subsection{The case of resonance $1:3$}

We have solved Eqs. (\ref{SING}) choosing $m=0.04$ $\left( \mu
=0.2\right) $ obtaining two singular solutions $X=\Omega ^{2}=9$, $Y=A^{2}=4$%
, and $f=F^{2}=0$, and $X=9.\,216\,088$, $Y=0.658\,061$, and $%
f=150.\,396\,416$. Both solutions correspond to isolated points.

\begin{figure}[h!]
\center
\includegraphics[width=10.5cm, height=7cm]{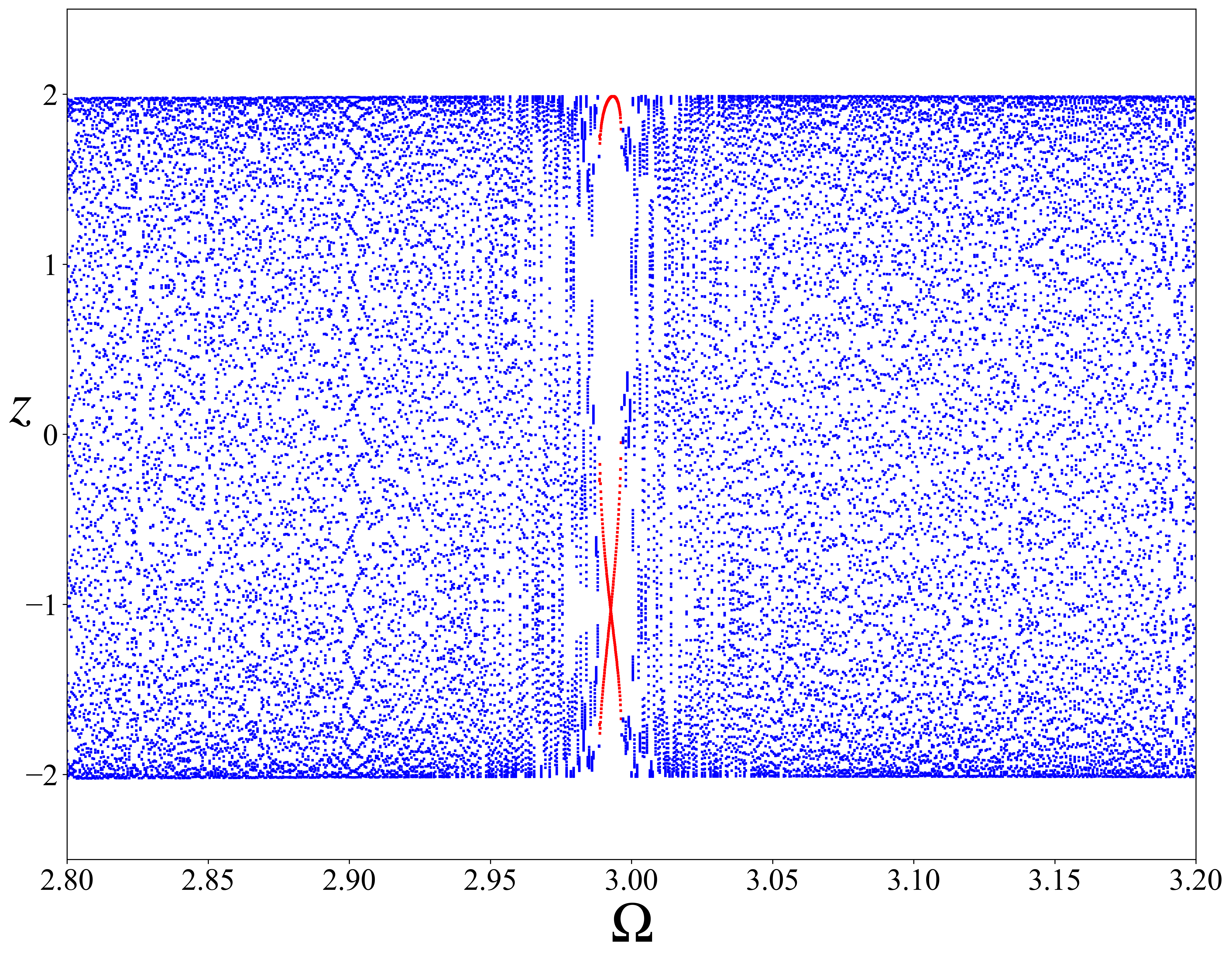}
\includegraphics[width=10.5cm, height=7cm]{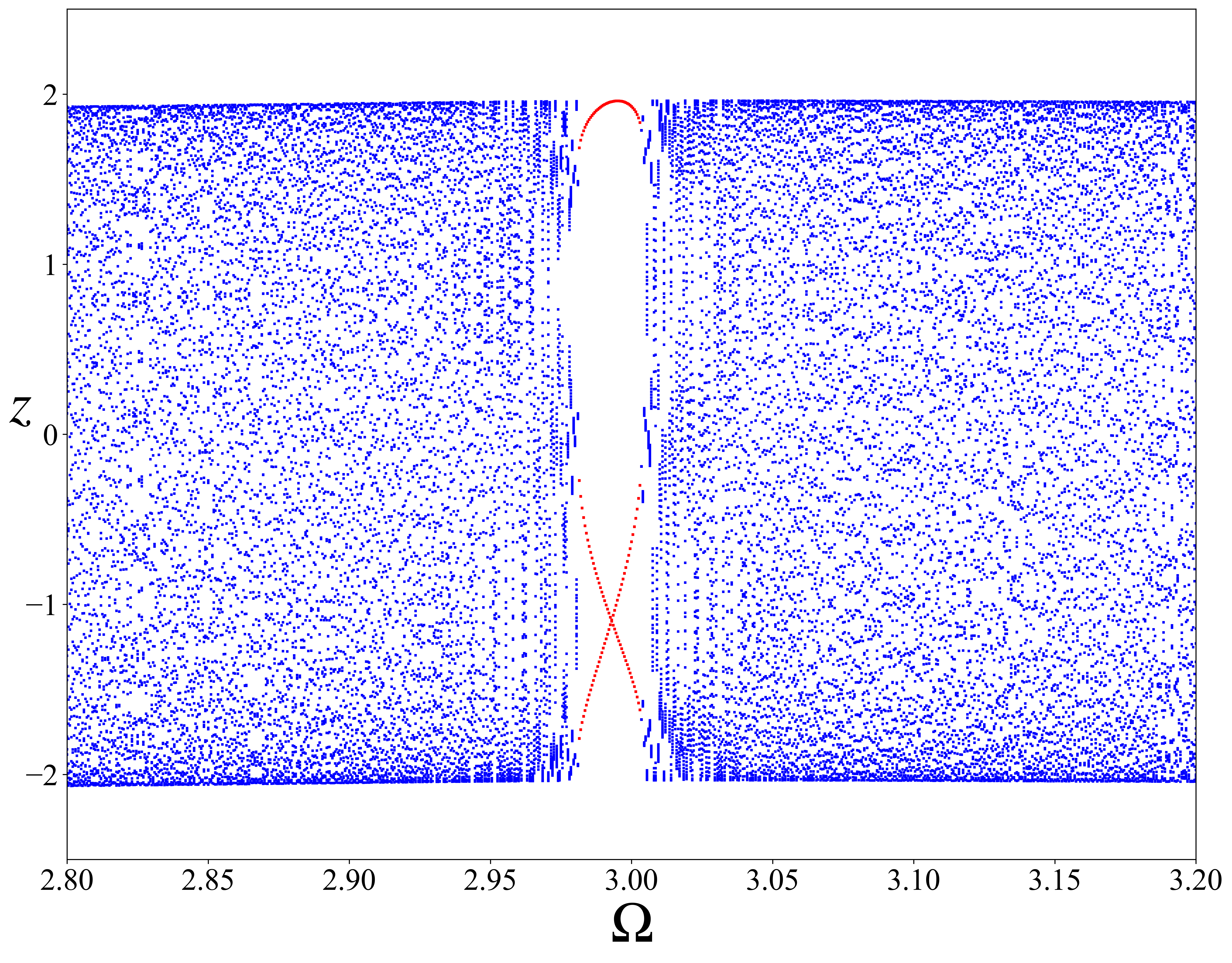}
\caption{The birth and growth of $1:3$ resonance (Red) -- numerical solutions of Eq. (%
\ref{vdP}). $\mu =0.2$; $F=0.2$ (top), and $F=0.6$ (bottom). Chaotic
dynamics is Blue.}
\label{F2}
\end{figure}

\begin{figure}[h!]
\center
\includegraphics[width=10.5cm, height=7cm]{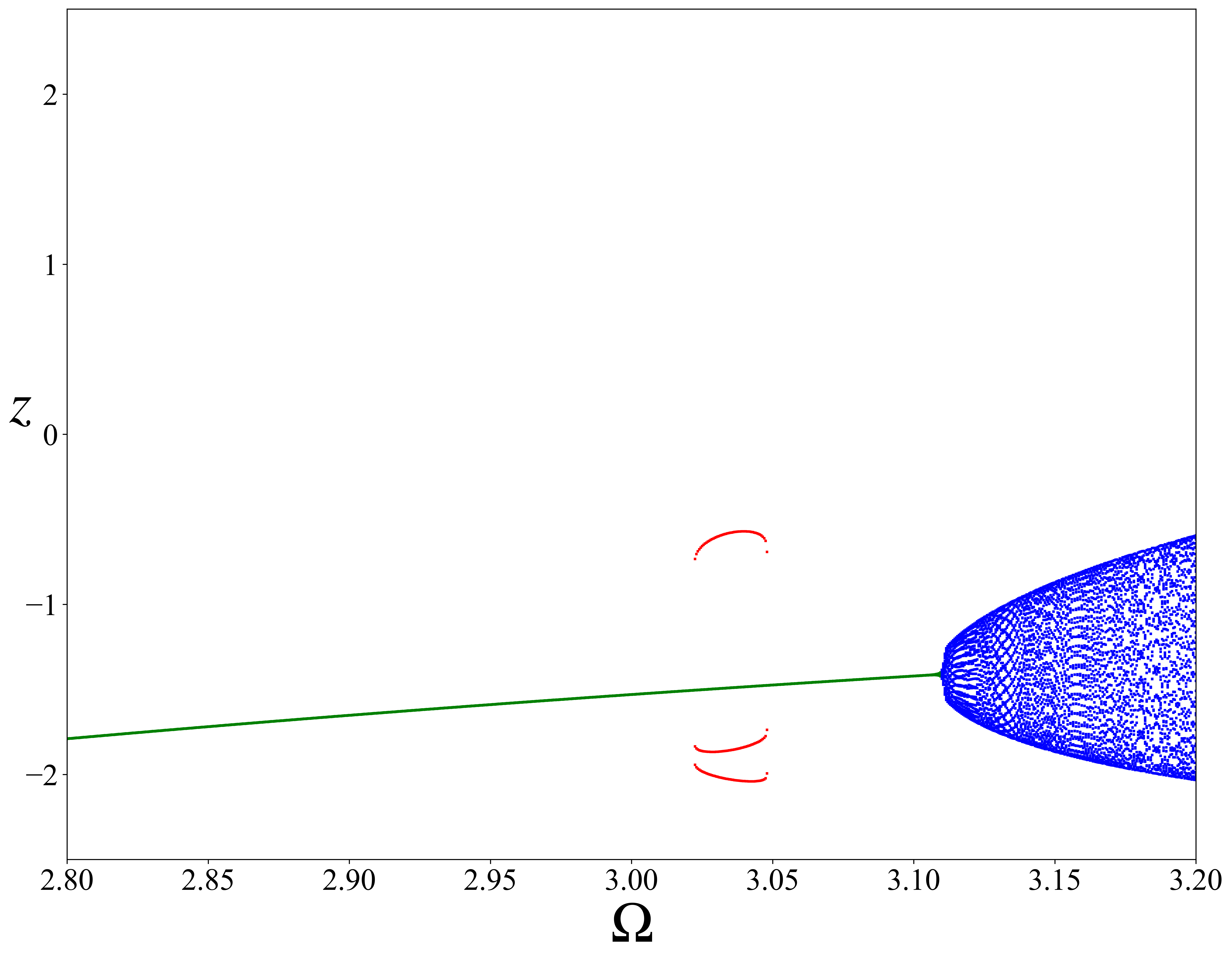}
\caption{The decay of $1:3$ resonance (Red) -- numerical solution of Eq. (\ref{vdP}). 
$\mu =0.2$; $F=12.25$. $1:1$ resonance is Green, chaotic dynamics is Blue. }
\label{F5}
\end{figure}

\begin{figure}[h!]
\center
\includegraphics[width=10.5cm, height=7cm]{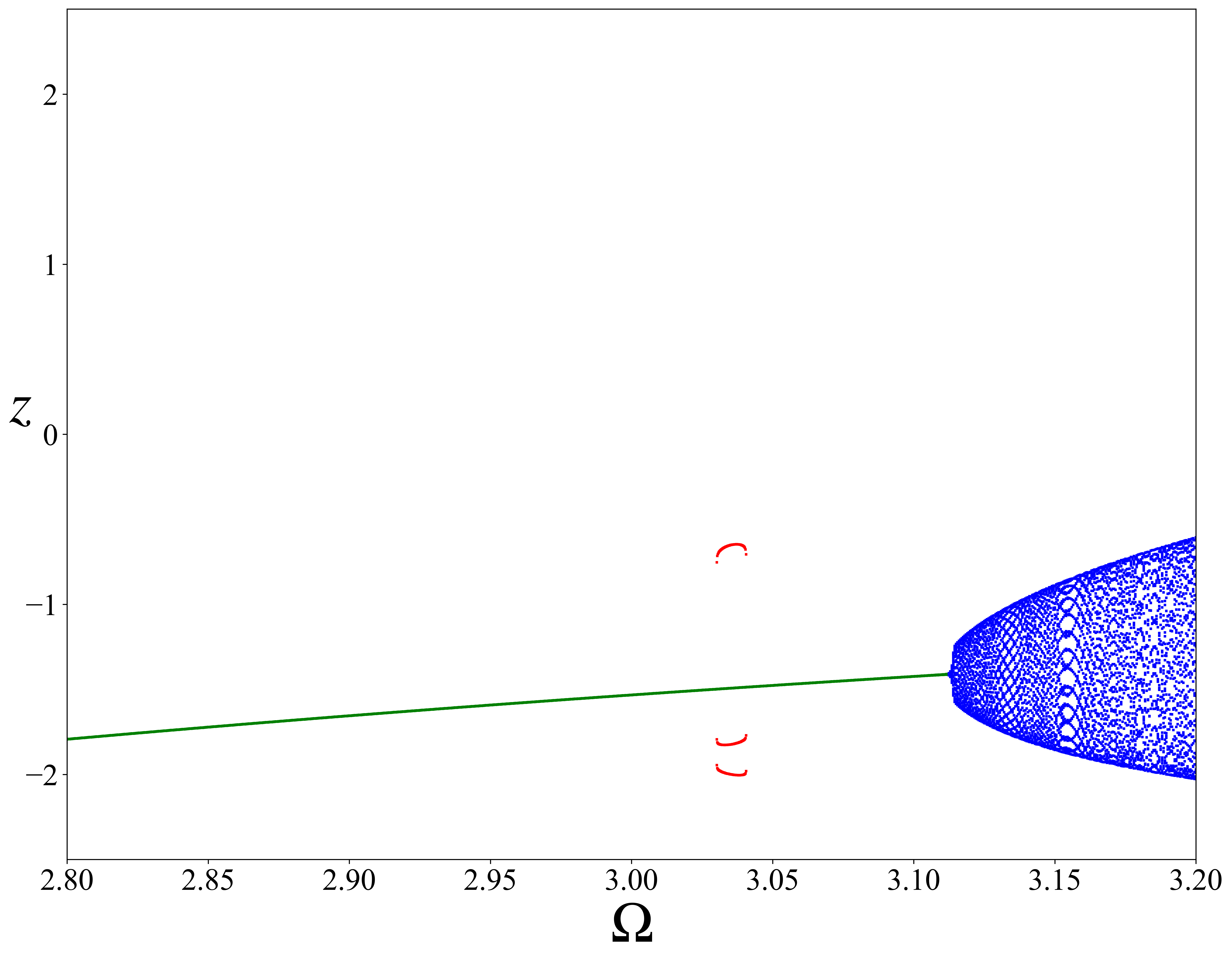}
\caption{The decay of $1:3$ resonance (Red) -- numerical solution of Eq. (\ref{vdP}). 
$\mu =0.2$; $F=12.27$. $1:1$ resonance is Green, chaotic dynamics is Blue.}
\label{F6}
\end{figure}

The first isolated point corresponds to the birth of $1:3$ resonance; see Figs. \ref{F2}, 
while the second isolated point corresponds to its decay; see Figs. \ref{F5} and \ref{F6}.

We have also solved Eqs. (\ref{VT}) choosing $m=0.04$ and $f=75.69$ $\left(
F=8.7\right) $ obtaining two vertical tangencies, $X_{1}=8.\,360\,693$, $%
Y_{1}=1.\,904\,477$ and $X_{2}=9.\,700\,773$, $Y_{2}=2.\,500\,266$; 
$\Omega _{1}=\sqrt{X_{1}}=2.\,891\,486$, $\Omega _{2}=\sqrt{X_{2}}%
=3.\,114\,606$. 

\begin{figure}[ht!]
\center
\includegraphics[width=10.5cm, height=7cm]{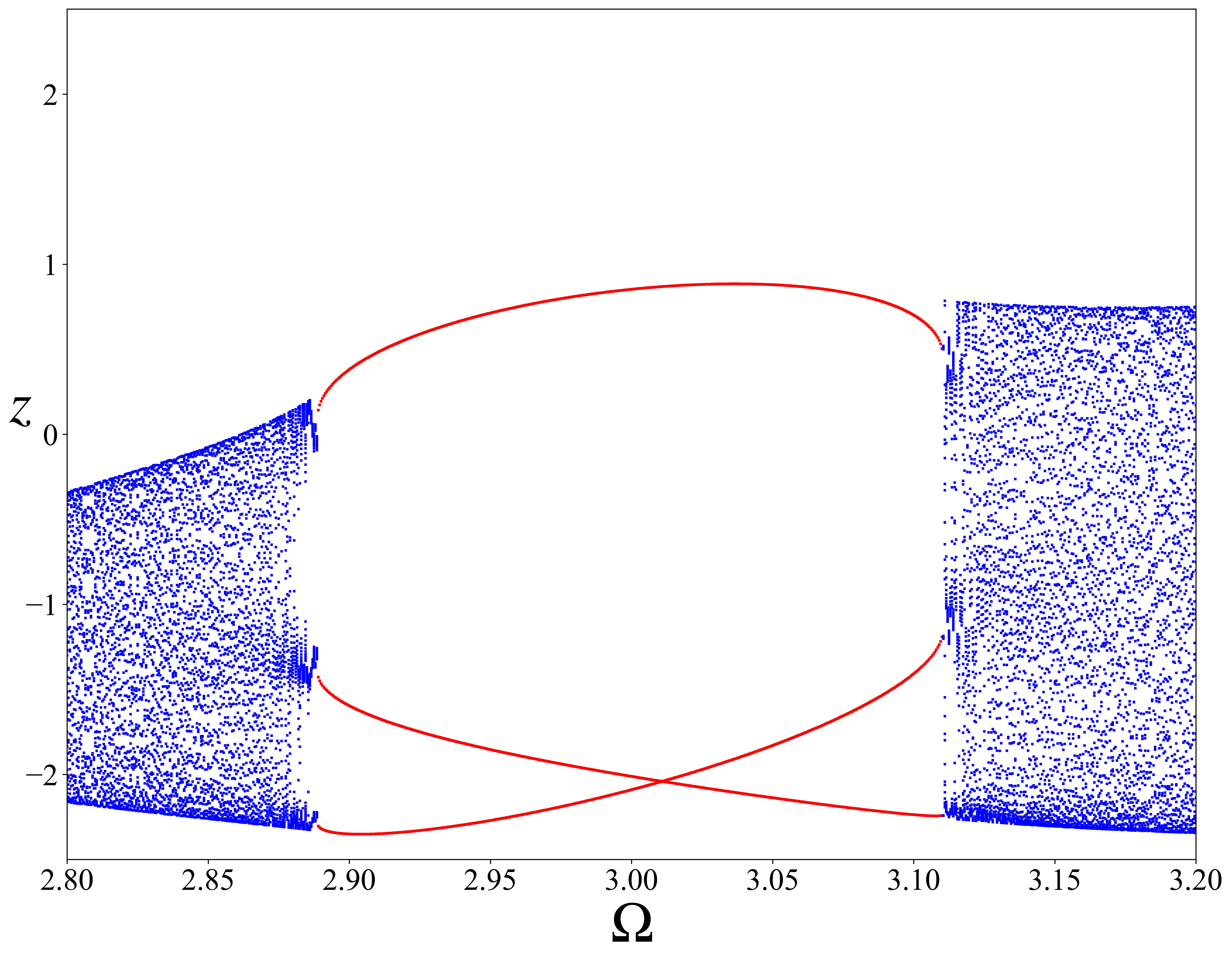}
\caption{The fully developed $1:3$ resonance (Red) -- numerical solution of Eq. (\ref%
{vdP}). $\mu =0.2$; $F=8.7$. Chaotic dynamics is Blue. }
\label{F4}
\end{figure}
Figure \ref{F4} shows a fully developed $1:3$ resonance. Positions of the
computed vertical tangencies, determining the beginning and the end of the
resonance, agree well with Fig. \ref{F4} -- the resonance is in the interval 
$\Omega \in \left( 2.8890,3.1100\right) $. 

Figures \ref{F5} and \ref{F6} show decay of the $1:3$ resonance. 
The resonance disappears completely for $F=12.28$, which agrees well 
with the computed value of $F$ for the second singular point, 
$F=\sqrt{150.\,396\,416}=12.\,263\,622$. 

\section{Summary}
In this work, we have studied the formation and growth of $1:1$ and $1:3$
resonances in the forced van der Pol oscillator, Eq. (\ref{vdP}). Our
approach consists of applying the geometry of algebraic curves to
investigate differential properties of amplitude-frequency (implicit)
response functions \cite{Gibson1998,Kyziol2022}. More precisely, we have
studied response functions for the $1:1$ and $1:3$ resonances; see Eqs. (\ref%
{1:1}) and (\ref{1:3}), respectively, computed singular points and vertical
tangencies of these functions, using Eqs. (\ref{SINGULAR}) (i.e., Eqs. (\ref%
{SING1:1}) or (\ref{SING})) as well as Eqs. (\ref{S1}) and (\ref{S3}) 
(i.e., Eqs. (\ref{VT1:1}) or (\ref{VT})). 

Isolated singular points correspond to the birth of a new
branch of solution, while vertical tangencies let us estimate the width of a
resonance. 

\medskip 

We have obtained the following results. 
\begin{enumerate}
\item In the case of the response function for the $1:1$ resonance, the
computed isolated point signals formation of the resonance, recovering a
well-known result, reasonably exact for small value of $\mu $ \cite%
{Jordan2007}. Moreover, the computed vertical tangencies permit the prediction of the
width of the growing resonance with good precision. 

\item In the case of the response function for the $1:3$ resonance, the
computed isolated points show the birth as well as decay of the resonance
with good precision. Moreover, the computed vertical tangencies provide a
good estimate of the width of the resonance.
\end{enumerate}

\section{Computational details}
\label{details} 
Nonlinear polynomial equations were solved numerically using
the computational engine Maple from Scientific WorkPlace 5.5. Figures \ref%
{Fa}, \ref{Fb}, \ref{Fc}, and \ref{Fd} were plotted with the computational
engine MuPAD from Scientific WorkPlace 5.5. Curves shown in bifurcation
diagrams in Figs. \ref{F1a}, \ref{F2}, \ref{F4}, \ref{F5}, and \ref{F6} were
computed and plotted running DYNAMICS, a program written by Helena E. Nusse and James A.
Yorke \cite{Nusse2012}, as well as our own programs, written in Pascal and Python \cite{Perez2007}.

\end{document}